\documentclass[runningheads]{llncs}
\usepackage{graphicx}
\usepackage{hyperref}
\usepackage{breakcites}
\hypersetup{
    colorlinks=true,
    linkcolor=blue,
    urlcolor=blue,
    citecolor=blue,
}
\begin{document}
\title{Detecting and Understanding Branching Frequency Changes in Process Models}
%
%
\author{Yang Lu\orcidID{0000-0002-9002-8650} \and
Qifan Chen\orcidID{0000-0003-1068-6408} \and
Simon Poon\orcidID{0000-0003-2726-9109}}
\authorrunning{Y. Lu et al.}
\titlerunning{Branching Frequency Changes in Process Models}
%
\institute{School of Computer Science, The University of Sydney, Sydney, NSW 2006, Australia 
\email{\{yalu8986, qche8411\}@uni.sydney.edu.au}\\
\email{simon.poon@sydney.edu.au}}
\maketitle              
\begin{abstract}
Business processes are continuously evolving in order to adapt to changes due to various factors. One type of process changes are branching frequency changes, which are related to changes in frequencies between different options when there is an exclusive choice. Existing methods either cannot detect such changes or cannot provide accurate and comprehensive results. In this paper, we propose a method which takes both event logs and process models as input and generates a choice sequence for each exclusive choice in the process model. The method then identifies change points based on the choice sequences. We evaluate our method on a real-life event log. Results show that our method can identify branching frequency changes in process models and provide comprehensive results to users. 

\keywords{Process Science \and Data Science \and Process drift \and Concept drift \and Branching frequency changes.}
\end{abstract}
\section{Introduction and Motivation}
Business processes are continuously changing in order to adapt changes to its execution environment, and such changes are called process drifts. Process drifts can be caused by different factors. Detecting and understanding such drifts can give us valuable information for process improvement.

One type of process drifts are changes of branching frequencies~\cite{Maaradji2017}, which is related to changes in frequencies between different options when there is an exclusive choice. Detecting and understanding branching frequency changes is important. For example, in a hospital triage process, if the proportion of emergency patients is found to be increasing, more resources can be sent to the emergency department to optimise the process.

\begin{figure}[h]
\includegraphics[width=\textwidth]{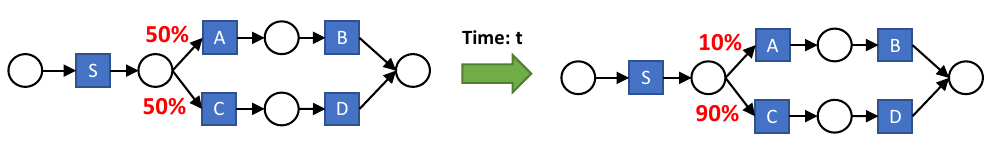}
\caption{An example branching frequency change from model 1 (left) to model 2 (right)} \label{example_1}
\end{figure}

Figure.\ref{example_1} shows an example branching frequency change. Before the change, the frequency of choosing A and C after S are equal, and after the change, the frequency of choosing C becomes much higher. Such a change can be detected by some process drift detection methods such as~\cite{Maaradji2015,Maaradji2017,Ostovar2016}. However, it is difficult to understand such changes even if they can be detected.

\begin{figure}[h]
\includegraphics[width=\textwidth]{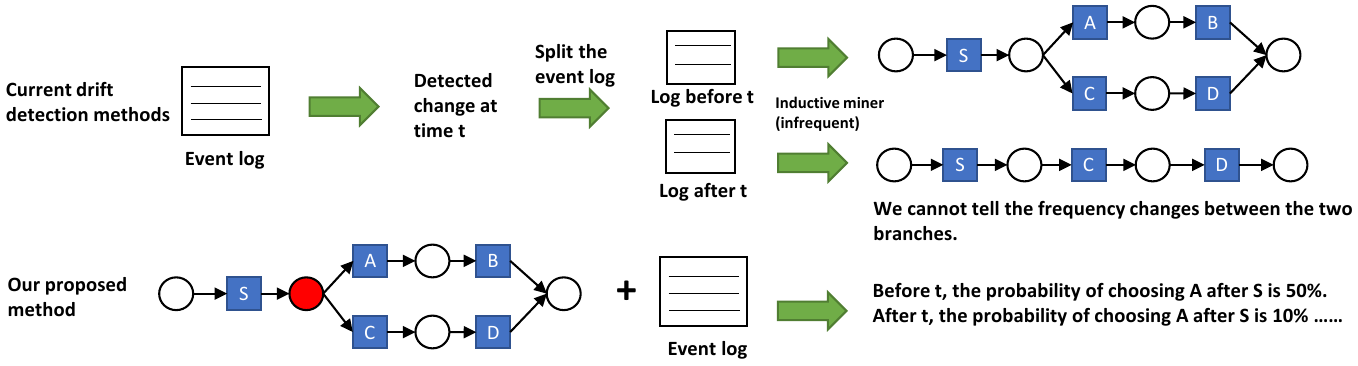}
\caption{Our method vs. current process drift detection methods when analysing the process drift in Fig.\ref{example_1}} \label{current_frequency_problem}
\end{figure}

Firstly, these process drift detection methods do not report if the returned process change points are related to process control-flow structure changes or branching frequency changes. A possible solution is to discover a process model based on sub-logs between every two consecutive detected process drift points, and if two consecutive process models are identical, we may conclude the detected process drift is related to branching frequency changes. However, the same process discovery algorithm can return different process models on the two sub-logs even if the process control-flow structure is unchanged. This is often due to the impacts of infrequent behaviours or the way for process discovery algorithms to deal with infrequent behaviours. We illustrate the problem in Fig.\ref{current_frequency_problem}. Secondly, most current process discovery algorithms can only return non-stochastic process models (i.e. they show all possible activity sequences, but they do not tell their probabilities). As a result, branching frequency changes cannot be reflected on discovered process models. Lastly, for real-life event logs, different models can be discovered when different process discovery algorithms are applied. Consequently, we need to obtain a process model before analyzing branching frequency changes.

In this paper, we narrow down the problem of branching frequency changes in process models to the changes of frequencies between different options when there is an exclusive choice. We propose a method which can detect such changes and provide comprehensive results for users. The rest of the paper is structured as follows: Section \ref{section2:2} is a literature review of related work. The proposed method is presented in Section \ref{section3:3}. Our method is then evaluated in Section \ref{section4:4}. We finally conclude our paper in Section \ref{section5:5}.

\section{Background}
\label{section2:2}
Recent studies in process drifts define a process drift as a time point when there is a significant change among process behaviours before/after the time point~\cite{Maaradji2015,Maaradji2017,Ostovar2016}. Based on the definition, \cite{Maaradji2015,Maaradji2017,Ostovar2016} use a sliding window to obtain two consecutive samples in the event log. Features are extracted from each sample. Then statistical hypothesis tests are performed among the consecutive samples, and if a significant change is found, a process drift can be reported. Those methods can successfully detect process control-flow structure changes and branching frequency changes, but are unable to produce a comprehensive results (i.e. they only return the time of process drifts). 

Based on the process drift points detected by \cite{Ostovar2016}, \cite{Ostovar2020} applies the inductive miner to get a process model between every two consecutive drift points and uses nature languages to describe the difference between discovered process models. \cite{Ostovar2020} can also report branching frequency changes in process models. However, as described in the previous section, we could get different process models even if the significant behaviour change is caused by a branching frequency change due to the impacts of infrequent behaviours or the ways for the process discovery algorithm  to deal with infrequent behaviours. In such a case, \cite{Ostovar2020} could mistakenly describe the process drift as a change in the process control-flow structure (See Fig.\ref{current_frequency_problem} for an example). In addition, \cite{Ostovar2020} can only be applied to block-structured \cite{Leemans2013} process models.

Some process drift detection methods are not based on statistical tests. For example, the TPCDD~\cite{zheng2017} reports a process drift whenever a new behaviour is observed or an existing behaviour is removed for a certain amount of time. As branching frequency changes will not bring new behaviours or remove existing behaviours to/from the process, TPCDD~\cite{zheng2017} cannot report such changes.

Other process drift detection methods aim at providing comprehensive results to users such as \cite{Yeshchenko2020,Yeshchenko2019,Seeliger2017,Stertz2018}. \cite{Seeliger2017} mines process models for different time periods and compares graph matrices of different models. \cite{Stertz2018} mines models for the first period of time and performs conformance checking on each new trace. A drift point is reported if there is a significant change on the conformance results. \cite{Yeshchenko2020,Yeshchenko2019} use Declare miners to represent the process, and a  comprehensive  visualisation  is  provided for users to understand process drifts. Those methods bring research about process drifts to a new stage, but still cannot provide comprehensive results for branching frequency changes in process models.

\section{The Proposed Method}
\label{section3:3}
\begin{figure}
\includegraphics[width=\textwidth]{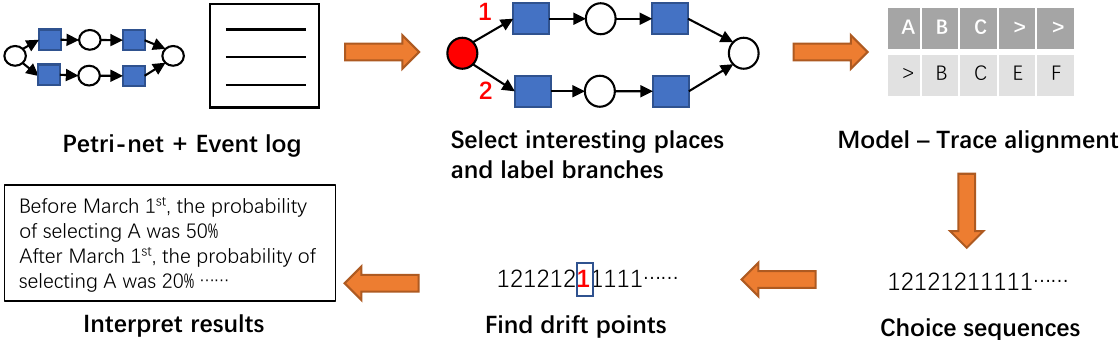}
\caption{An overview of the proposed method} \label{method_overview}
\end{figure}

Fig.\ref{method_overview} shows an overview of our proposed method. The inputs of the method are a Petri net and event log. Based on the inputs, the proposed method will detect branching frequency change points and generate a comprehensive report. More specifically, only sound Petri nets can be used as the input. For the definition of soundness, we refer to \cite{wmp2016}.
\subsection{Step 1: Discovering Process Models and Selecting Interesting Places}
In the first step, users should discover a process model based on the event log. The process model can be discovered by any process discovery algorithms or manually created by domain experts. In the method proposed in this paper, all process models should be transformed to sound Petri nets.

In a sound Petri net, an exclusive choice split is caused by a place with more than one outgoing arcs. If users are interested in frequency changes in one or multiple exclusive choices, they can select such places and only focus on process drifts on the places they select. Otherwise all places with more than one outgoing arcs can be selected. In this paper, we call selected places ``interesting places''.

\subsection{Step 2: Discovering Choice Sequences}
\begin{figure}
\includegraphics[width=\textwidth]{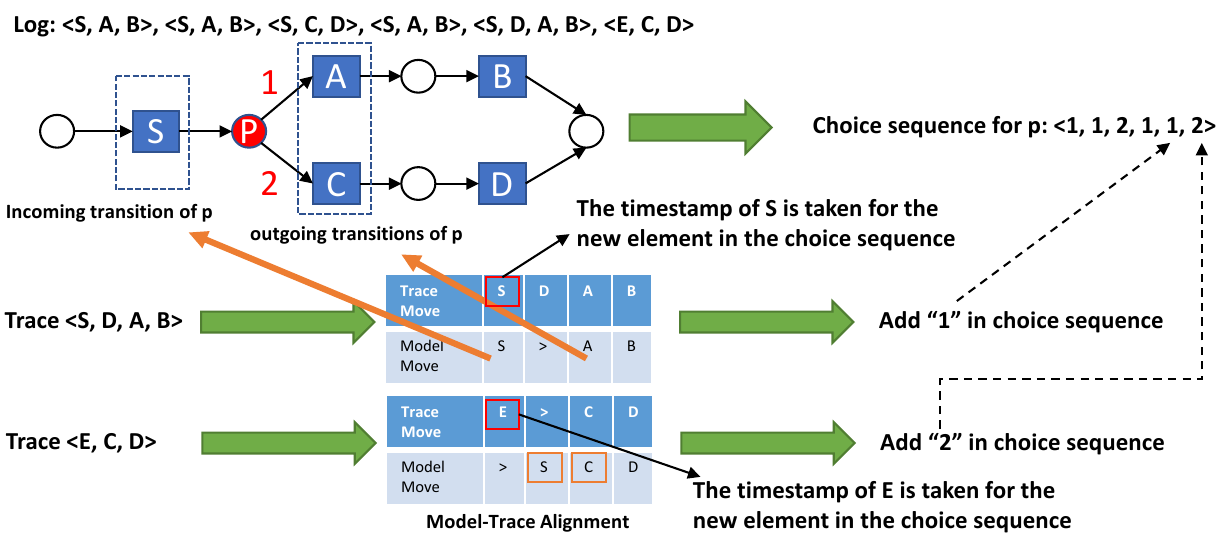}
\caption{Discovering Choice Sequences} \label{choice_sequence}
\end{figure}

To describe the sequence of decisions each time a choice is needed to be made on the process model, we propose a new concept named choice sequence. A choice sequence for a place p is a sequence of categorical variables, each refers to an outgoing arc of p and corresponds to a timestamp. We firstly give a label to each outgoing arc of an interesting place. Then each time a choice is made, a new element is added into the choice sequence. In our implementation, we use an integer to label each arc. Fig.\ref{choice_sequence} shows a simple example. For place p, the first outgoing arc is labeled as 1 while the second arc is labeled as 2. Each time the first (second) arc is selected, 1 (2) is added into the choice sequence. 

However, when dealing with real-life event logs and process models, some traces may not conform with the process model. In addition, Petri nets can also contain hidden transitions (i.e transitions without activity labels.). Those factors make it hard for us to determine which outgoing arc of the interesting place has been selected during process executions. To accurately produce choice sequences, we perform model-trace alignments~\cite{wmp2012} between the event log and process model. For each trace, we find its corresponding model move sequence. Then for each interesting place p, we find its incoming and outgoing transitions. If we find two consecutive model moves where the first one belongs to the incoming transitions of p, and the second model move belongs to the outgoing transitions of p, we add a new element into the choice sequence of place p. In addition, we take the timestamp of the closest trace move which is before (or corresponded to) the first model move for the new element. We finally sort the choice sequence based on the timestamps. Fig.\ref{choice_sequence} shows a complete example of this step.

\subsection{Step 3: Detecting Change Points}
After choice sequences are detected, the problem of finding branching frequency changes is changed into identifying a change point on sequences of categorical variables. We can then apply other change point detection methods to identify change points. In our implementation, we use an external library called ruptures~\cite{Truong2020}. Ruptures is a python package which contains a collection of change point detection algorithms for various dynamic systems~\cite{Truong2020}. It has also been used in detecting process drifts in~\cite{Yeshchenko2020,Yeshchenko2019}.

Since each place has its own choice sequence, each time a change point is discovered, we can easily know what has been changed. For example, in Fig.\ref{choice_sequence}, if we detect a change point on the choice sequence of place p, we know the process drift is related to the choice between activity A and C after S. As a result, by calculating the frequency of ``1" and ``2" in the choice sequence, a comprehensive report about the frequency changes between ``A" and ``C" after ``S" can be provided.

\section{Evaluation}
\label{section4:4}
The proposed method is implemented as a Python program based on the PM4PY \cite{berti2019} framework, and a prototype is publicly available\footnote{\url{https://github.com/bearlu1996/FrequencyChange}}. 

We evaluate our method on a real-life event log which is publicly available from the ``4TU Data Center"\footnote{\url{https://doi.org/10.4121/uuid:0c60edf1-6f83-4e75-9367-4c63b3e9d5bb}}. The event log describes a ticketing management process of the help desk of an Italian software company. 

The event log has also been used to evaluate the process drift characterization method in \cite{Ostovar2020}. \cite{Ostovar2020} reports two process drifts in the event log, one of which is related to a branching frequency change. Although we do not know the exact ground truth about process drifts in the event log, we can validate our results by comparing to the results provided by \cite{Ostovar2020}.

\subsection{Step 1: Discovering Process Models and Selecting Interesting Places}
\begin{figure}
\includegraphics[width=\textwidth]{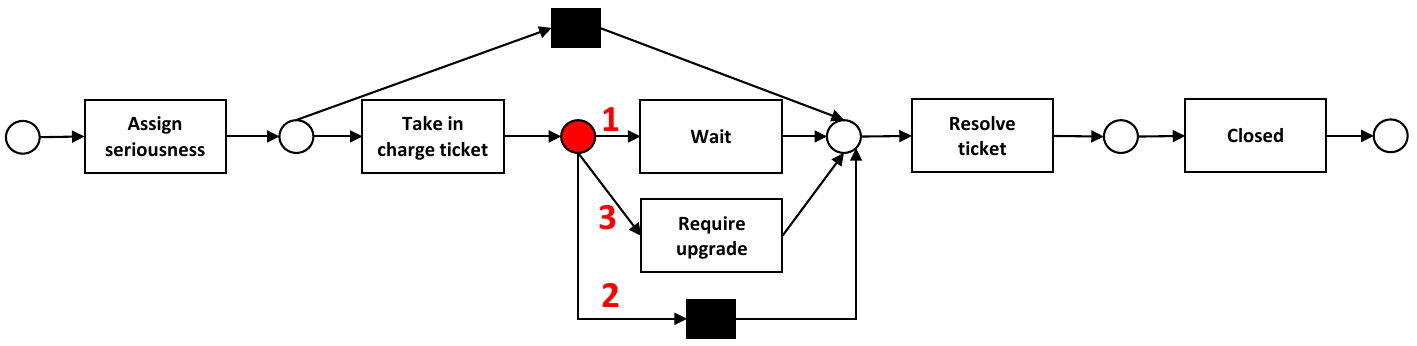}
\caption{Process model of the ``Italian help desk" log from \cite{Ostovar2020}} \label{desk_model}
\end{figure}

In order to compare our results with \cite{Ostovar2020}, we directly use the process model discovered by \cite{Ostovar2020} as the input model for our method. The model is discovered by the inductive miner, and infrequent behaviours are filtered out. The model is presented in Fig.\ref{desk_model}. In this study, we select the place before ``Wait" as the interesting place (marked in red) since we want to learn the change of frequencies between choosing ``Wait", ``Require upgrade" and ``Resolve ticket" after ``Take in charge ticket". Then the three outgoing arcs are automatically labeled as ``1", ``2" and ``3". Each label represents an outgoing arc after ``Take in charge ticket". 

\subsection{Step 2: Discovering Choice Sequences}
After the interesting place is set, its incoming and outgoing transitions are then identified. In our case, the incoming transition of the selected place is ``Take in charge ticket", while the outgoing transitions are ``Wait", ``Require upgrade" and a hidden transition. We then perform model-trace alignments and generate a choice sequence for the selected place. Finally, we sort the choice sequence by the timestamp of each element.

\subsection{Step 3: Detecting Change Points}
\begin{figure}
\includegraphics[width=\textwidth]{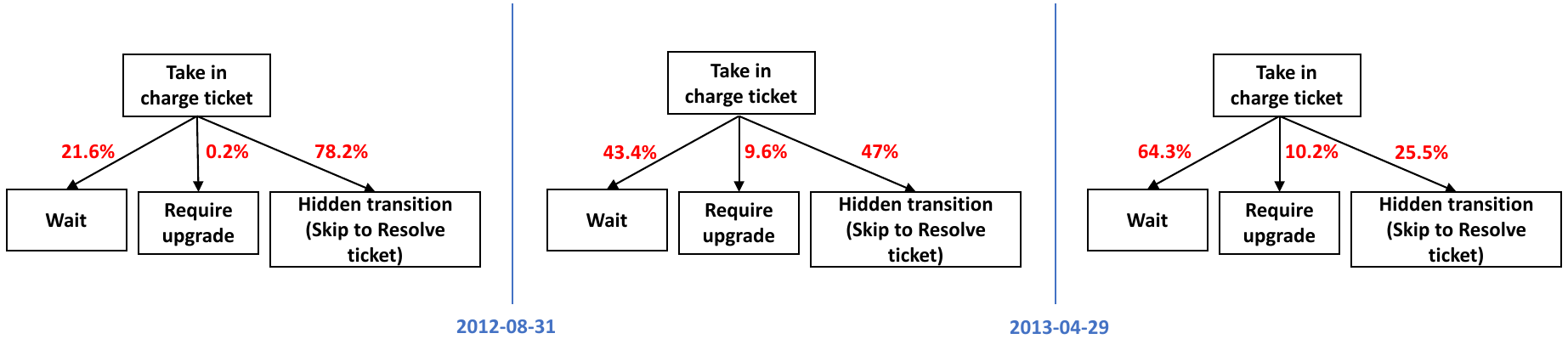}
\caption{Visualisation of the detected branching frequency changes} \label{desk_results}
\end{figure}

Finally, the choice sequence is treated as input for Ruptures~\cite{Truong2020} to identify change points\footnote{In this study, we use Pelt as the search method, CostRbf as the cost function. The penalty value is set to 5. For details, please refer to \cite{Truong2020}.}. Two change points are detected in total, the first one is on Aug 31st, 2012, and the second one is on Apr 29th, 2013. A comprehensive result is presented in Fig.\ref{desk_results}. Before Aug 31st, 2012, The frequency of choosing ``Require upgrade" after ``Take in charge ticket" is very low (0.2\%). The frequency of skipping to ``Resolve ticket" (78.2\%) is much higher than ``Wait" (21.6\%). Between Aug 31st, 2012 and Apr 29th, 2013, the frequency of choosing ``Require upgrade" is much higher than before (9.6\%), and the frequency of ``wait" and skipping to ``Resolve ticket" are almost the same (around 45\%). After the second drift on Apr 29th, 2013, the frequency of choosing ``Require upgrade" almost remains unchanged, but the frequency of choosing ``Wait" (64.3\%) is much higher than the frequency for choosing skipping to ``Resolve ticket" (25.5\%).

The first drift point is detected by both our method and \cite{Ostovar2020}\footnote{In \cite{Ostovar2020}, the drift is detected on Sep 11th, 2012.}. In addition, our method also reports one more frequency change point. Since the frequency change between ``Wait" and skipping to ``Resolve to ticket" is significant (the change is around 20\%), we believe the second drift point we find is correct.

\section{Conclusion}
\label{section5:5}
In this paper, we present a method to detect branching frequency changes in process models. The method can not only detect changes in processes, but can also provide comprehensive results to users. By retrieving information about branching frequency changes from processes, subsequent analysis can be performed for process improvement.

Future work may include different aspects. Firstly, we can integrate our method with a user interface which can visualize branching frequency changes in process models. Secondly, we aim at applying other change point detection methods to detect changes on choice sequences and empirically compare different methods.

%
%
%

\begin{thebibliography}{8}
\bibitem{Maaradji2015}
Maaradji, A., Dumas, M., Rosa, M., Ostovar, A.: Fast and accurate business process drift detection. In: Motahari-Nezhad, H.R., Recker, J., Weidlich, M. (eds.) BPM 2015. LNCS, vol. 9253, pp. 406–422. Springer, Heidelberg (2015)

\bibitem{Maaradji2017}
Maaradji, A., Dumas, M., La Rosa, M., Ostovar, A.: Detecting sudden and gradual drifts in business processes from execution traces. IEEE TKDE 29(10), 2140–2154 (2017)

\bibitem{Ostovar2016}
Ostovar, A., Maaradji, A., La Rosa, M., ter Hofstede, A.H.M., van Dongen, B.F.V.: Detecting drift from event streams of unpredictable business processes. In: Comyn-Wattiau, I., Tanaka, K., Song, I.-Y., Yamamoto, S., Saeki, M. (eds.) ER 2016. LNCS, vol. 9974, pp. 330–346. Springer, Cham (2016)

\bibitem{wmp2012}
van der Aalst, W.M.P., Adriansyah, A., van Dongen, B.F.: Replaying history on process models for conformance checking and performance analysis. Wiley Interdiscip. Rev. Data Min. Knowl. Discov. 2(2), 182–192 (2012)

\bibitem{Truong2020}
Truong, C., Oudre, L., Vayatis, N.: Selective review of offline change point detection methods. Signal Process. 167 (2020)

\bibitem{Yeshchenko2020}
Yeshchenko, A., Di Ciccio, C., Mendling, J., Polyvyanyy, A.: Visual Drift Detection for Event Sequence Data of Business Processes. arXiv preprint (2020)

\bibitem{Yeshchenko2019}
Yeshchenko, A., Di Ciccio, C., Mendling, J., Polyvyanyy, A.: Comprehensive pro-
cess drift detection with visual analytics. In: Laender, A.H.F., Pernici, B., Lim,
E.-P., de Oliveira, J.P.M. (eds.) ER 2019. LNCS, vol. 11788, pp. 119-135. Springer,
Cham (2019)

\bibitem{Ostovar2020}
Ostovar, A., Leemans, S.J. and Rosa, M.L.: Robust drift characterization from event streams of business processes. ACM Transactions on Knowledge Discovery from Data (TKDD), 14(3), pp.1-57 (2020)

\bibitem{Leemans2013}
Leemans, S.J.J., Fahland, D., van der Aalst, W.M.P.: Discovering block-structured process models from event logs - a constructive approach. In: Colom, J.-M., Desel, J. (eds.) PETRI NETS 2013. LNCS, vol. 7927, pp. 311–329. Springer, Heidelberg (2013)

\bibitem{zheng2017}
Zheng, C., Wen, L., Wang, J.: Detecting process concept drifts from event logs. In:
OTM CoopIS, pp. 524–542 (2017)

\bibitem{Seeliger2017}
Seeliger, A., Nolle, T., Mühlhäuser, M.: Detecting concept drift in processes using graph metrics on process graphs. In: Proceedings of the 9th Conference on Subject-Oriented Business Process Management, p. 6:1 (2017)

\bibitem{Stertz2018}
Stertz, F., Rinderle-Ma, S.: Process histories - detecting and representing concept drifts based on event streams. In: Panetto, H., Debruyne, C., Proper, H.A., Ardagna, C.A., Roman, D., Meersman, R. (eds.) OTM 2018. LNCS, vol. 11229, pp. 318–335. Springer, Cham (2018)

\bibitem{wmp2016}
van der Aalst, W.M.P.: Process Mining - Data Science in Action. Springer, Heidelberg
(2016)

\bibitem{berti2019}
Berti, A., van Zelst, S.J., van der Aalst, W.M.P.: Process mining for python (PM4PY): bridging the gap between process - and data science. CoRR abs/1905.06169 (2019)
\end{thebibliography}
%

\end{document}